\documentclass{emulateapj}
\slugcomment{Draft Version, Accepted for Publication in ApJ}
\usepackage{times}
\usepackage{bm}% bold math
\usepackage{color}

\newcommand{\bole}{{\bm  e}}

\newcommand{\bolE}{{\bm  E}}

\newcommand{\bolB}{{\bm  B}}
\newcommand{\bolJ}{{\bm  J}}
\newcommand{\bolV}{{\bm  V}}

\newcommand{\beq}{\begin{equation}}
\newcommand{\eeq}{\end{equation}}
\newcommand{\beqn}{\begin{eqnarray}}
\newcommand{\eeqn}{\end{eqnarray}}
\newcommand{\beqno}{\begin{equation*}}
\newcommand{\eeqno}{\end{equation*}}
\newcommand{\beqnno}{\begin{eqnarray*}}
\newcommand{\eeqnno}{\end{eqnarray*}}
\newcommand{\etal}{{et al. }}

\shorttitle{Parabolic Structure of the M87 Jet}
\shortauthors{Nakamura \& Asada}

\begin{document}

\title{The Parabolic Jet Structure in M87 as a Magnetohydrodynamic Nozzle}

\author{Masanori Nakamura\altaffilmark{1} and Keiichi Asada\altaffilmark{1}}
\affil{
$^1$Institute of Astronomy \& Astrophysics, Academia Sinica, P.O. Box
23-141, Taipei 10617, Taiwan \\
{\tt nakamura@asiaa.sinica.edu.tw, asada@asiaa.sinica.edu.tw}}

\begin{abstract}
The structure and dynamics of the M87 jet from sub-milli-arcsec to
arcsecond scales are continuously examined. We analysed the VLBA
archival data taken at 43 and 86 GHz to measure the size of VLBI
cores. Millimeter/sub-mm VLBI cores are considered as innermost jet
emissions, which has been originally suggested by Blandford \&
K\"{o}nigl. Those components fairly follow an extrapolated parabolic
streamline in our previous study so that the jet has a single power-law
structure with nearly five orders of magnitude in the distance starting
from the vicinity of the supermassive black hole (SMBH), less than 10
Schwarzschild radius ($r_{\rm s}$). We further inspect the jet parabolic
structure as a counterpart of the magnetohydrodynamic (MHD) nozzle in
order to identify the property of a bulk acceleration.  We interpret
that the parabolic jet consists of Poynting-flux dominated flows,
powered by large amplitude, nonlinear torsional Alfv\'en waves.  We
examine the non-relativistic MHD nozzle equation in a parabolic
shape. The nature of trans-fast magnetosonic flow is similar to the one
of transonic solution of Parker's hydrodynamic solar wind; the jet
becomes super-escape as well as super-fast magnetosonic at around $\sim
10^{3}\,r_{\rm s}$, while the upstream trans-Alfv\'enic flow speed
increases linearly as a function of the distance at $\sim 10^{2}$ --
$10^{3}\, r_{\rm s}$. We here point out that this is the first evidence
to identify these features in astrophysical jets. We propose that the
M87 jet is magnetically accelerated, but thermally confined by the
stratified ISM inside the sphere of gravitational influence of the SMBH
potential, which may be a norm in AGN jets.
\end{abstract}


\section{Introduction}
\label{sec: Intro}
How does the acceleration and collimation take place as well as the
formation process in astrophysical jets? Although electromagnetic and/or
magnetohydrodynamic (MHD) mechanisms are frequently invoked to extract
energy and momentum from the compact object and/or the accretion disk
\citep[{\em e.g.},][]{BZ77, BP82, PN83, US85, SU86, L87, HN89, L91,
PP92, M99, M01}, none of real astrophysical jet systems has yet been
examined to confirm these theoretical properties in the quantitative
manner.

M87 is one of the nearest active galaxies \citep[16.7 Mpc; ][]{J05} that
exhibit relativistic outflows. The mass of the central supermassive
black hole (SMBH) $M_{\bullet}$ is measured with a range $3.2 \times
10^{9} M_{\sun}$ \citep[{\em e.g.},][]{M97} to $6.6 \times 10^{9}
M_{\sun}$ \citep[]{G11}. This largest mass gives an apparent angular
size $\sim$ 8 $\mu$as for the Schwarzschild radius $r_{\rm s} \equiv 2 G
M_{\bullet}/c^2$. This galaxy therefore provides a unique opportunity to
study the relativistic outflow with the highest angular resolution in
units of $r_{\rm s}$. Extended synchrotron emissions of one-sided jet,
emerging from the nucleus, have been well-studied at multi-wavelengths
from radio to X-ray bands \citep[]{R89, O89, B95, P99, B99, J99, M02,
WY02, PW05, H06, L07, KOV07}, which cover a spatial scale $\sim 0.4$ mas
-- 14 arcsec with the viewing angle 14$^{\circ}$ \citep[]{WZ09},
corresponding to $2 \times 10^2 - 7 \times 10^6 \ r_{\rm s}$ in
de-projection.

Very recently, there are several attempts to examine the inner jet
structures on M87; i) \citet[]{H11} performed the core shift measurement
by using multi-frequency, phase referencing VLBA observations,
indicating that the 43 GHz VLBI core is located at $\sim 20 \, r_{\rm
s}$ from the central engine (presumably, the SMBH and/or accretion
disk). They conclude that the measured frequency dependence of the core
shift $\propto \nu^{-0.94 \pm 0.09}$ is in good agreement with a
synchrotron self-absorbed ``conical'' jet \citep[]{BK79, K81}.  ii)
\citet[][hereafter AN12]{AN12} investigated the structure of the M87
jet, from milliarcsec (mas) to arcsec scales by utilizing the images
taken with EVN, MERLIN, and VLBA. AN12 reveal that the jet maintains a
``parabolic'' shape $z \propto r^{1.73 \pm 0.05}$ in a de-projected
distance of $10^{2}$ -- $10^5 \, r_{\rm s}$, where $r$ is the radius of
the jet emission and $z$ is the axial distance from the core. iii)
\citet[]{D12} conducted the Event Horizon Telescope (EHT) observation at
a wavelength of 1.3 millimeter (mm), deriving the size of 230 GHz VLBI
core to be $5.5 \pm 0.4 \, r_{\rm s}$. This is smaller than the diameter
for the innermost stable circular orbit (ISCO) of a retrograde accretion
disk, suggesting that the M87 jet may be powered by a prograde accretion
disk around a spinning SMBH.

In the standard picture of VLBI jets, the compact radio core (VLBI core)
is widely believed as the throat of a diverging (with a constant opening
angle), conical jet \citep[][]{BK79}. Alternatively, \citet[][]{DM88}
proposed that the VLBI core is identified as the first re-collimation
Mach disk - oblique shock system. The VLBI core at mm wavelength is
located to be $10^{4-6} \, r_{\rm s}$ where the synchrotron emission is
completely optically thin [where the optical depth $\tau (\nu)$ is
fallen to $\sim 1$], which corresponds to the turnover in the spectrum
\citep[]{J07}. The jet acceleration and collimation zone is speculated
in the upstream of the mm-wave radio core. In the present paper, we
discuss about the jet acceleration and collimation zone in M87 by
introducing a parabolic jet streamline as a context of MHD nozzle, based
on resent results (i) - (iii) listed above, as well as our own VLBA data
reductions at 43/86 GHz. We apply the standard picture of the VLBI core
(as an innermost jet at giving frequency) to M87, one of the most
resolved AGN jets, to inspect a feasibility of the physical
interpretation of the core.

The paper is organized as follows. In \S \ref{sec: Proper}, we give a
review of observed proper motions in M87 and their interpretation.  \S
\ref{sec: MHD-Jets} summarizes a general picture of MHD jets and points
to be confirmed in M87. We outline observation details in \S \ref{sec:
Obs}, followed by our results in \S \ref{sec: Res}. We analyze the jet
parabolic streamline as a counterpart of the MHD nozzle in \S \ref{sec:
Analysis}. Discussions and summary are given in \S \ref{sec: Dis} and \S
\ref{sec: Sum}. Throughout the paper, we use the de-projected distance
along the jet (assuming the viewing angle 14$^{\circ}$) from the central
SMBH in units of $r_{\rm s}$. We note that 1 mas corresponds to 517
$r_{\rm s}$ in de-projection.

\section{Proper Motions Associated to the Flow Speed}
\label{sec: Proper}

In a general framework on asymptotic evolutions of MHD jets in parabolic
streamlines, both bulk acceleration and collimation take place
simultaneously. In the present paper, we investigate the jet
acceleration in order to confirm the jet acceleration and collimation
zone of M87. One of the useful tools to study the jet kinematics is the
direct measurement of a proper motion (an ``apparent'' speed $V_{\rm
app}$ of moving components) by the multi-epoch observations. Let
$\theta_{\rm v}$ be a viewing angle between our line of sight and an
``intrinsic'' speed $V_{\rm int}$ along the jet. Then $V_{\rm app}$ is
corrected in the units of the speed of light $c$ ($\beta=V/c$) through
the usual relationship:
\beqn
\label{eq: V_app-V_int}
\beta_{\rm app}=\frac{\beta_{\rm int} \sin \theta_{\rm v}}{1-\beta_{\rm int}
\cos \theta_{\rm v}}.
\eeqn

\begin{figure}
\begin{center}
\includegraphics[scale=0.45, angle=-90]{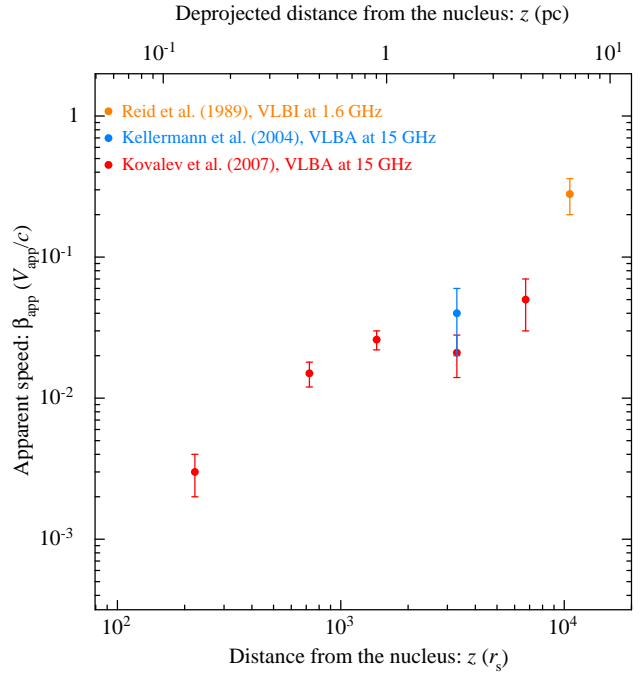} \caption{
\label{fig: proper} 
Distribution of the apparent speed $\beta_{\rm app} (= V_{\rm app}/c)$
as a function of the projected distance from the core. Measurements by
VLBA at 15 GHz \citep[red and blue,][respectively]{KOV07, K04}, and
global VLBI at 1.6 GHz \citep[orange,][]{R89}.}
\end{center}
\end{figure}

Proper motions of the M87 jet in the parabolic region ($\lesssim
10^{5}\, r_{\rm s}$) have been probed by VLBI observations during the
past two decades \citep[{\em e.g.},][]{R89, K04, KOV07} as shown in
Fig. \ref{fig: proper}, exhibiting a series of subluminal motions. Note
that a peak of apparent motions, that is superluminal $\beta_{\rm app}
\sim 6$, is located at around the HST-1 complex \citep[]{B99}, where the
jet has a structural transition from the parabolic to conical streamline
as shown in AN12.  As is clearly seen in Fig. \ref{fig: proper}, it is
remarkable that observed apparent speed keeps increasing by almost two
orders of magnitude at $z = 10^{2} - 10^{4} \,r_{\rm s}$, where the jet
structure is being parabolic.

In the literature as listed above, these subluminal motions upstream of
HST-1 have been generally interpreted as stationary patterns as standing
shocks and/or some modes of plasma instabilities. This may be because
one could consider the M87 jet structure as a cylindrical or conical
structure \citep[{\em e.g.,}\rm][]{KOV07}. When the supersonic,
hydrodynamic jet, which is highly over-pressured against the ambient
ISM, is causally disconnected from its driving engine, forming
re-collimation shock due to the pressure imbalance between the inside
and outside the jet \citep[]{S83}. Or perhaps, growing plasma
instability and its related turbulence may be speculated. However, we
suggest {\em neither} would be the case in M87. As AN12 revealed, the
jet structure is parabolic so that the jet outer edge still has a causal
contact with its interior, where the jet could be magnetically
accelerated up to over $10^{5}\, r_{\rm s}$ scale.

\citet[]{L09} investigate the parse-scale AGN jet kinematics in the
complete MOJAVE survey; a median rms dispersion of apparent speeds in
the overall sample is three times larger than that value within an
individual jet. Two-sided jets in relatively low-luminosity and nearby
radio galaxies provide notable evidence that apparent motions in the jet
and counter jet are correlated. These survey may give an additional
insight that there is a characteristic flow that describes each jet;
apparent motions are not simply (shock/instability) patterns, which
propagate slowly and/or relatively stable, but do indeed represent
underlying bulk flows. Even $V_{\rm app}$ is corrected by a viewing
angle with Eq. (\ref{eq: V_app-V_int}), random distribution of $V_{\rm
pat}$ at two sides should be generally expected no matter whether the
jet and counter jet are intrinsically symmetric (similar kinematics of
each side due to a common origin of the jet initiation). In at least NGC
1052 and Cygnus A, \citet[]{L09} confirm a one-to-one association
between features on the two sides, supporting a hypothesis that observed
proper motions reflect the real jet flow. It would seem far more natural
to have $V_{\rm int}$ tied to the jet (fluid) bulk speed $V$
\citep[]{B95} as
\beqn
\label{eq: V_int-V}
\beta_{\rm int} \lesssim \beta.
\eeqn

Another important remark is that MHD jets, which are powered by
torsional Alfv\'en waves (TAWs) \citep[]{US85, SU85}, are intrinsically
robust compared to purely hydrodynamic jets regarding nonlinear wave
perturbations. When wave amplitudes are no longer small, most
disturbances such as a (magneto-)sonic wave cease to propagate as a wave
with a constant profile, but instead distort in time (steepening into a
shock). However, Alfv\'en waves can continue to propagate without
distortions even when their amplitudes are large
\citep[]{P81}. Recently, this unusual, but very interesting property is
also verified in the relativistic Alfv\'enic perturbations
\citep[]{HLM12}. We thus suggest that observed subluminal patterns in
M87 may correspond to slowly, but presumably non-relativistic fluids
that are magnetically accelerated along the jet parabolic streamline. We
will examine this in \S \ref{sec: Analysis}.

\section{General Picture of MHD Jets in Global Scales}
\label{sec: MHD-Jets}

We here remark one of pioneering studies in the field of MHD outflows by
\citet[][hereafter BP82]{BP82}. Authors examined axisymmetric, steady,
self-similar cold magnetohydrodynamic outflows from the Keplerian
accretion disk, showing that not only the initiation of bulk outflow
takes place by a ``magneto-centrifugal'' effect (without any support by
gas pressure), but also continued bulk acceleration as well as
simultaneous self-collimation can be expected in the super-Aflv\'enic
regime by dominant toroidal (azimuthal) magnetic fields. The flow
eventually becomes super fast magnetosonic [note that this ``classical''
fast point (CFP) is {\em not} a critical point in self-similar
solutions] and a causal contact between the jet interior and its outer
edges by magnetosonic waves will be maintained until the flow will pass
the ``modified'' fast point (MFP), where the ``collimation'' component
of the poloidal speed $V_{p} \sin \theta$ toward the jet rotating axis
(where $\theta$ is the local angle between the poloidal magnetic field
and any straight line drawn from the origin in spherical geometry)
becomes comparable to the local fast magnetosonic speed.

The concept of the MFP has been recognized by BP82, but their MHD
outflow never passes the MFP (even at large distance from the disk
surface). First steady, self-similar, non-relativistic MHD (NRMHD)
outflow solutions, which pass the MFP, are introduced by
\citet[]{VTST00}. Very recently, \citet[]{PMM10} successively solve
steady, self-similar, relativistic MHD (RMHD) flow solutions beyond the
MFP. Their extensive survey of the parameter space reveals the location
of the MFP spans from $10^{4} \, r_{\rm g}$ to $10^{8} \, r_{\rm g}$,
where $r_{\rm g} \equiv G M_{\bullet}/c^2$ is the gravitational radius;
the location of the MFP may be sensitive to the poloidal spherical angle
of the Alfv\'en point (AP), $\theta_{\rm A}$ (a distance $D_{\rm MFP}$
monotonically decreases from $> 10^{6} \, r_{\rm g}$ to $\sim 10^{4} \,
r_{\rm g}$ while $\theta_{\rm A}$ increases as $40^{\circ} \rightarrow
60^{\circ}$) and so-called the magnetization parameter $\sigma_{\rm M}$;
Poynting-to-matter energy flux ratio ($D_{\rm MFP}$ increases from $\sim
10^{5} \, r_{\rm g}$ to $\sim 10^{8} \, r_{\rm g}$ if $\sigma_{\rm M}$
increases as $\sim 1 \rightarrow 60$) \citep[]{PMM13}. Both
non-relativistic and relativistic solutions exhibit {\em an asymptotic
bulk acceleration until the flow reaches the MFP}, where the streamline
remains parabolic \cite[{\em see also}\rm][but, note that an MFP would
exist at infinity in their solutions]{LCB92, VK03a}.

In the standard process of NRMHD jet acceleration, the poloidal jet
speed exceeds the local escape speed around trans-Alfv\'enic regime
\citep[]{KS97, R97, VTST00}. Even in the RMHD jet, this feature may be
held; \citet[]{PMM13}'s very recent steady, axisymmetric, {\em
semi}-self-similar formulation {\em first passes through all three
singular points} [the ``modified'' slow point (MSP), AP, and MFP]. In a
relativistic framework, the gravity does {\em not} follow a
self-similarity and thus none of past semi-analytical RMHD solutions in
a {\em full}-self-similar manner could not obtain a complete solution to
pass all singular points. In their fiducial solution, the location of
the MSP falls within the range of a surface layer of the radiatively
inefficient accretion flows (RIAFs) \citep [$H \sim r$ and $B_{\phi}
\sim B_{p}$, where $H$ the disk scale height, $B_{\phi}$ the toroidal
component, and $B_{p}$ the poloidal magnetic field components,
respectively: {\em e.g.},][]{M01}. A nominal magneto-centrifugal process
does take place after passing the MSP and the outflow immediately passes
the AP without becoming relativistic ($\gamma = 1$, where $\gamma$ is
the Lorentz factor). Around the AP, the flow kinetic energy ($\gamma-1$)
overtakes the gravitational potential energy. Beyond the AP, the
magnetic acceleration becomes effective, increasing the poloidal flow
speed towards the relativistic regime, whereas ($\gamma-1$) only becomes
relativistic further downstream along the jet.

In this paper, we suggest observed subluminal proper motions on 0.1 to
10 pc scales (Fig. \ref{fig: proper}) may exhibit slowly, but certainly
accelerated fluid components, which can be described in the framework of
NRMHD parabolic jets. \citet[]{M06} discussed general relativistic MHD
(GRMHD) simulation results, as the ergosphere-driven jet
\citep[][]{BZ77} does not fit into the observed feature in M87.
Instead, he suggests that the broader emission component in the M87 jet
could be due to the slow disk wind, which interacts with the surrounding
ISM. We also suggest that the outermost part of the jet (even it may
still be powered by the frame-dragging in the ergosphere), the flow
speed remains non-relativistic $\gamma \simeq 1$ at $\sim 5 \times
10^3\,r_{\rm g}$ although the inner part becomes relativistic $\gamma
\sim 5$ at this scale. The dominant toroidal magnetic pressure at the
outermost part of the jet has a pressure-matching with the ambient
non-magnetized coronal (thermal) pressure shown Fig. 9 in \citet[]{M06}.

Another theoretical effort by \citet[]{G09} dealt with a steady,
axisymmetric, multi-layer outflows (relativistic spine jet and
non-relativistic sheath wind), nicely reproducing such observed
properties of M87 as an asymptotic collimation and edge-brightened
emission. Also, \citet[]{K07, K09} performed axisymmetric RMHD numerical
simulations to examine the magnetic jet acceleration. Jet outer boundary
is assumed to be confined by a rigid wall of adopted conical/parabolic
shapes, while the poloidal fluxes inside the jet are differentially
collimated toward the central axis. We are especially interested in
their ``differential rotation'' model; they prescribe a non-uniform
rotation to drive the jet at the base. $\gamma$ increases mostly in the
intermediate region between the inner spine and outer sheath, on the
other hand $\gamma$ remains almost unity in the outer sheath region even
at the further downstream of $10^{4}$ -- $10^{6}$ scales of the jet
launching ($\sim$ a radius of the event horizon).

In general, (G)RMHD jets may be covered by the slowly moving outer
sheath which determines the global shape of jets under the pressure
equilibrium with the ISM. Therefore, we suggest the non-relativistic
bulk flow in M87 may be relevant to the edge-brightened sheath part of
the jet although the inner spine part of the jet has not yet been
clearly identified. As we will argue in \S \ref{sec: Dis-offset}, a
future VLBI imaging of the jet emission higher than 43 GHz will be a
crucial test whether the spine/sheath structure exists.

\section{Archival Data and Data Reduction}
\label{sec: Obs}

We analysed archival data (BW088 and BJ045) of VLBA observations at 43
and 86 GHz. VLBA observation at 43 GHz was carried out on 2007 January
27 at 43.127 and 43.135 GHz using all ten stations of the VLBA, but Kitt
Peak station. VLBA observation at 86 GHz was carried out on 2007 January
17 at 86.253 and 86.269 GHz using all stations of the VLBA, but Saint
Croix and Hancock stations. Each IF has 16-MHz bandwidth in both
observations.  The data were correlated at the VLBA correlator at
Socorro.

An {\it a priori} amplitude calibration for each station was derived
from the system temperatures measured during observations and antenna
gain curve information from standard files. Opacity correction was made
during amplitude calibration process using the AIPS task APCAL. Fringe
fitting was performed on each IF and polarization independently using
the AIPS task FRING. After delay and rate solutions were determined, the
data were averaged over 12 seconds in each IF and self-calibrated using
Difmap. We note that the VLBA observations were conducted with dual
polarization, and we independently reduced the data at 86 GHz at both
polarizations and confirmed consistency of the obtained images.

\begin{figure*}[bt]
\centerline{{\includegraphics[scale=1.02, angle=0]{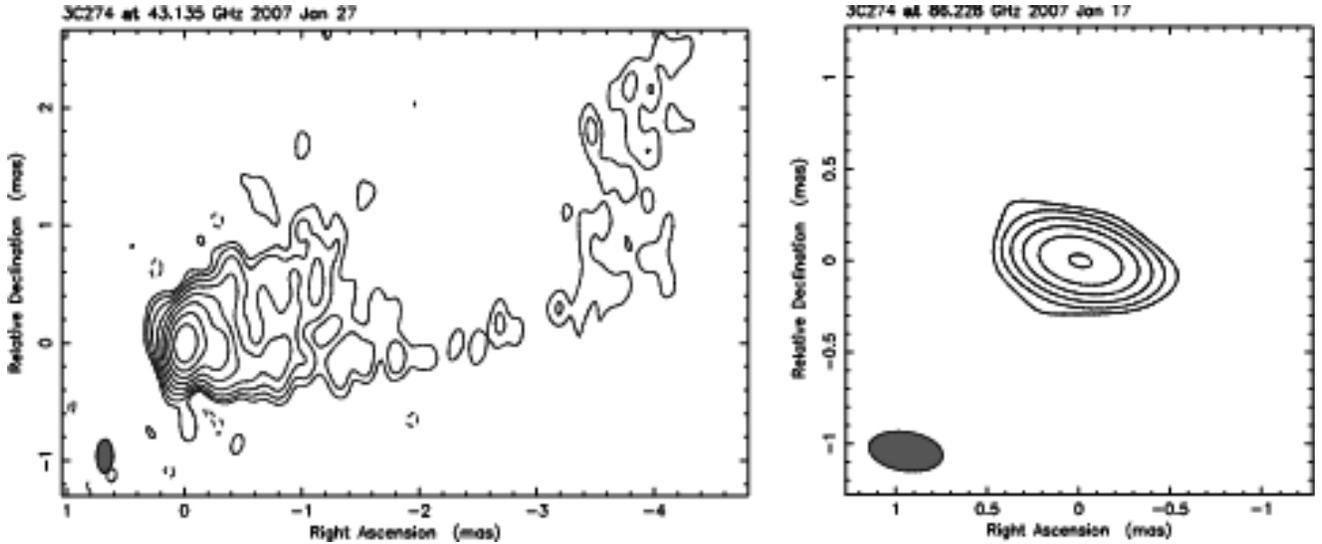}}} \caption{
\label{fig:image} 
VLBA image of M87 at 43 GHz ({\em left}). Contours are plotted at -1,
1, 2, ..., 1024 $\times$ 1.9 mJy beam$^{-1}$, which is three-times the
residual r.m.s. noise.  The synthesized beam is 0.281 mas $\times$ 0.139
mas with the major axis at a position angle of -0.07$^{\circ}$. VLBA
image of M87 at 86 GHz ({\em right}).  Contours are plotted at -1, 1,
2, ..., 1024 $\times$ 26.3 mJy beam$^{-1}$, which is three-times the
residual r.m.s. noise.  The synthesized beam is 0.406 mas $\times$ 0.212
mas with the major axis at a position angle of 81.6$^{\circ}$.}
\end{figure*}

\section{Results}
\label{sec: Res}

We show the VLBA images at 43 and 86 GHz in Fig. \ref{fig:image}. Image
qualities are summarized in Table 1. The bright core emission at the
eastern edge of the jet and continuous jet emission are detected in the
image at 43 GHz. The jet shows an edge-brightened structure as seen in
the previous images \citep[][]{J99, KOV07, L07, H11}. On the other hand,
we only see the bright core emission at 86 GHz, while the jet emission
is not detected. This is because of the less image sensitivity, since
large telescopes are not included in our imaging. Note that previous
GMVA observations show some diffuse emission of the jet at this
frequency \citep[]{K06}.

We define the core with a Gaussian model that was fitted to the
innermost bright region in all the images. The measured size of the VLBI
core at 43 (core$_{43}$) and 86 GHz (core$_{86}$) were 182 $\pm$ 53 and
100 $\pm$ 55 $\mu$as, corresponding to $23 \pm 7 \, r_{\rm s}$ and $13
\pm 7 \, r_{\rm s}$, respectively. The measured size of core$_{86}$ is
in good agreement with 99 $\pm$ 21 $\mu$as in the previous GMVA
observation \citep[]{L08}.  Also we remark the size of the VLBI core at
230 GHz (core$_{230}$) is derived by the EHT observation as $5.5 \pm 0.4
\, r_{\rm s}$ in the correlated flux density analysis by circular
Gaussian models \citep[]{D12}.

\begin{figure*}
\centerline{{\includegraphics[scale=1.4, angle=0]{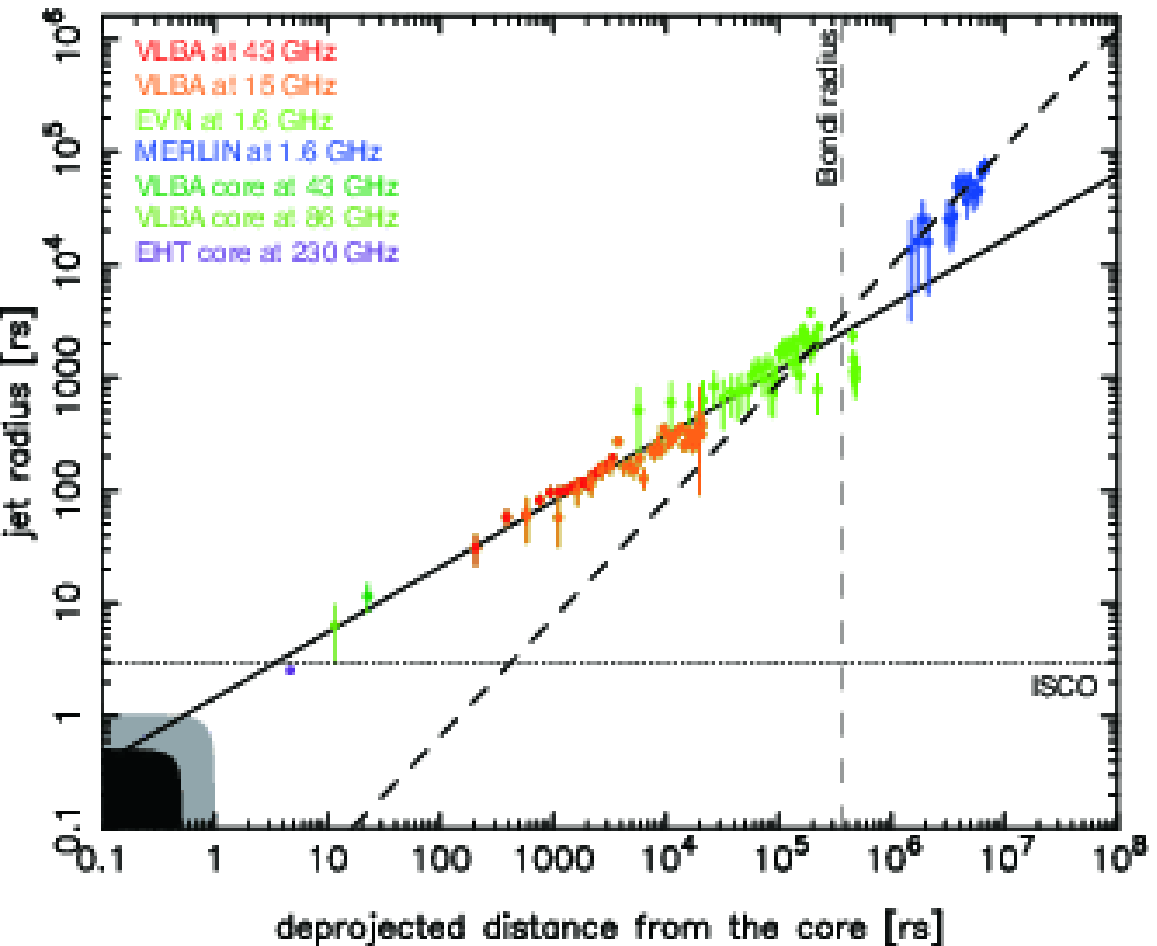}}}
\caption{\label{fig: r-z} 
Distribution of the radius of the jet as a function of the de-projected
distance from the core in units of $r_{\rm s}$. Readers can refer to
Fig. 1 in \citet[]{AN12} for detailed description. Three data points of
VLBI cores (at 43, 86, and 230 GHz) are added as the most inner jet
emissions at each frequency. The solid line is the linear least-square
for data points except three inner cores (VLBA at 43/86 GHz and EHT at
230 GHz), indicating the parabolic streamline with a power-law index $a$
of $1.73 \pm 0.05$. On the other hand, the dashed line indicates the
conical streamline with $a$ of $0.96 \pm 0.1$.  HST-1 is located around
$5 \times 10^{5}\,r_{\rm s}$. The thin dashed line denotes the Bondi
accretion radius located at $R_{\rm B} \simeq 3.8 \times 10^{5}\,
r_{\rm s}$. The black area shows the size of the minor axis of the event
horizon of the spinning black hole with maximum spin. The gray area
indicates the size of the major axis of the event horizon of the
spinning black hole with maximum spin, and corresponds to the size of
the event horizon of the Schwarzschild black hole. The thin dotted line
indicates the size of the inner stable circular orbit (ISCO) of the
accretion disk for the Schwarzschild black hole.
}
\end{figure*}

Figure \ref{fig: r-z} shows the radius of the jet as a function of the
de-projected distance from the central SMBH, which is essentially
similar to Fig. 2 in AN12, but additional three data points of VLBI
cores (core$_{43,\, 86,\, 230}$) are included. Current innermost
emissions of the M87 jet are lied at $\sim$ a few of $10^2 \, r_{\rm s}$
from the central SMBH, which have been observed by 43 GHz VLBA
observations. However, if we consider a hypothesis, which has been
introduced by \citet[]{BK79}, that the VLBI core is the most upstream of
the optical thin synchrotron emission at a given frequency \citep[thus a
position of the ``self-absorbed core'' depends on an observed frequency:
{\rm e.g.},][]{K81}, we can examine the further innermost jet structure
by using VLBI cores with 43 GHz and higher frequencies.

In order to determine an offset position in the jet axial direction
(from the core position to the central engine), we follow the formula in
\citet[]{H11}. An absolute offset $\Delta z$ can be determined as a
function of the frequency as
\beqn \label{eq: core-shift}
\Delta z (\nu/{\rm GHz})=A \nu^{-\alpha} \ ({\rm mas}), 
\eeqn
where $\alpha=0.94 \pm 0.09$ and $A=1.40 \pm 0.16$. We therefore
estimate $\Delta z$ (43 GHz) = $21.2 \pm 6.2 \, r_{\rm s}$, $\Delta z$
(86 GHz) = $10.9 \pm 4.7 \, r_{\rm s}$, and $\Delta z$ (230 GHz) = $4.34
\pm 2.17 \, r_{\rm s}$, respectively.  As is shown in Fig. \ref{fig:
r-z}, three points of VLBI cores (core$_{43}$, core$_{86}$, and
core$_{230}$) are plotted under our assumption that the axial
offset position at infinite frequency $\Delta_{\infty} (\rightarrow$\,0)
is the location where the SMBH and/or accretion disk plane exist. Note
that VLBA core sizes at 5.0 and 8.4 GHz coincide the jet width derived
in VLBA observations at 43 GHz (Hada et al., in private communications),
thus it may be reasonable to interpret frequency-dependent VLBI core as
an innermost synchrotron emission where the jet emissions become
optically thin.

The parabolic jet seems to follow a single power-law streamline with
nearly five orders of magnitude in distance.  Three points of VLBI cores
are not used for a fit to data points to derive a power-law index $a$ of
$1.73 \pm 0.05$ in AN12. One, {\em however}, has to bear in mind the
following quest; {\em where does jet origin exist and how does its
non-thermal emission is initiated at some higher frequency}? It is
nevertheless useful for us to examine the nature of the jet parabolic
structure; how it is maintained under the stratified ISM in the dominant
gravitational potential by the central SMBH.

In the following sections, based on the MHD jet theory, we analyze the
bulk acceleration of the trans-Alfv\'enic flow and derive the
approximate MHD nozzle equation of the trans-fast magnetosonic jet in \S
\ref{sec: Analysis}. We discuss about the formation of the parabolic
streamline in M87 as well as a potential limit of exploring the
innermost jet emission by using VLBI core shift measurements as we quest
above in \S \ref{sec: Dis}.

\section{Analysis of the Parabolic Streamline}
\label{sec: Analysis}
\subsection{Bulk Acceleration of the Trans-Alfv\'enic Flow}
\label{sec: TAW}

In order to inspect the property of the NRMHD jet acceleration in the
trans-Alfv\'enic regime, let us follow the wave dynamics of TAWs
propagating along a global poloidal magnetic field, which is originally
argued by \citet[]{SU85}. We consider a poloidal magnetic flux tube
given by
\beqn
B_{p} \Sigma = \Psi = {\rm constant},
\eeqn
where $\Sigma$ is a cross section of the flux tube and $\Psi$ is the
magnetic flux. $\Sigma$ is assumed to be $\Sigma \propto r^{2}$.  Under
the WKB approximation, the conservation of the energy flux density of
the TAW is considered through a circular (cross-sectional) area around
the rotation axis in cylindrical coordinate ($r, \phi, z$) as
\beqn
\label{eq: WKB}
\frac{\rho \delta V_{\phi}^2 V_{\rm A}}{B_{p}}={\rm constant},
\eeqn
here, $\rho$, $\delta V_{\phi}$ and $V_{\rm A} \equiv B_{p}/(4 \pi
\rho)^{1/2}$ are the gas density, azimuthal component of the velocity
field (a transverse displacement against the poloidal magnetic line of
force drives a TAW), and Alfv\'en speed, respectively. From this
equation, the following relation can be obtained:
\beqn
\label{eq: Vf-rho}
\delta V_{\phi} \propto \rho^{-1/4}.
\eeqn

Perturbed quantities in the transverse direction satisfy
\beqn
\delta V_{\phi} = - \frac{\delta B_{\phi}}{\sqrt{4 \pi \rho}},
\eeqn
which is another property of the shear Alfv\'en wave \citep[]{P81}. Thus
TAWs can be characterized as an equality of ratios between poloidal and
toroidal components in the magnetic field,
\beqn
\label{eq: TAW}
\frac{\delta B_{\phi}}{B_{p}} = - \frac{\delta V_{\phi}}{V_{\rm A}}. 
\eeqn
Note that the wave amplitudes ($\delta B_{\phi}$ and $\delta V_{\phi}$)
are not necessarily small, but can be arbitrary finite. In a linear
regime ($|\delta B_{\phi}|$ $\ll$ $B_{p}$), a perturbation of the TAW is
incompressible [only magnetic tension force $({\bolB}_{p} \cdot
\nabla)B_{\phi}$ induces the incompressible Alfv\'en mode], while if a
wave amplitude is not negligibly small $|\delta B_{\phi}|/B_{p} \gtrsim
O(1)$, the TAW also drives the compressional Alfv\'en mode. In a steady,
axisymmetric MHD solution of BP82-type cold MHD outflow, $|\delta
B_{\phi}| \simeq B_{p}$ is expected in the trans-Alfv\'enic flow $V_{p}
\approx V_{\rm A}$ \citep[{\em e.g.},][]{KP00}. On the other hand, in a
non-cold MHD outflow, the initial bulk acceleration takes place by the
thermal pressure gradient force and then magnetic acceleration processes
take over beyond the MSP. In general, steady, axisymmetric solutions
exhibit $|\delta B_{\phi}| \gtrsim B_{p}$ at around Alfv\'enic point in
both NRMHD \citep[]{KS97, VTST00} and RMHD \citep[]{PMM13} regimes.
Hereafter $\delta B_{\phi}$ is replaced by $B_{\phi}$ in the paper.

We therefore remark large amplitude, nonlinear TAWs  play an
important role in the trans-Alfv\'enic jet dynamics, which has been
first examined in full two-dimensional MHD simulations \citep[]{US85,
SU85, SU86}. A magneto-centrifugal process of BP82 by quasi-linear TAWs
is no longer effective, while, the toroidal magnetic pressure gradient
force in the poloidal direction $\nabla_{p} (B_{\phi}^2/8\pi)$ becomes
dominant in the trans-Alfv\'enic regime. The uncoiling of $B_{\phi}$
produces a compression of the plasma in the poloidal direction, which
drives fast magnetosonic waves, powering the bulk acceleration. On the
other hand, the hoop stress: $-B_{\phi}^2/4 \pi r$ squeezes the plasma
toward the central axis. Physical picture described above is named as
``sweeping magnetic twist'' mechanism by \citet[]{US86}.

Considering the trans-Alfv\'enic NRMHD bulk flow powered by nonlinear TAWs,
order-of-magnitude estimations of the one-dimensional ($z$) equation of
motion give
\beqn 
\label{eq: motion} 
V_{z} \sim \frac{B_{\phi}^2}{4 \pi \rho} V_{\rm A}^{-1}. 
\eeqn
By combining eqs. (\ref{eq: Vf-rho}), (\ref{eq:
TAW}), and (\ref{eq: motion}), \citet[]{SU85} derive
\beqn 
V_{z} \sim B_{p}^{-1} = {\rm constant}.
\eeqn
Thus, the terminal velocity of TAW-driven MHD jet model with an
initially uniform axial field ($B_{p}=$ constant) is a constant with an
order of trans-Alfv\'enic speed \citep[{\em e.g.},][]{SU85, KMS98,
KS05}.

On the other hand, the terminal velocity of TAW-driven MHD jet with a
non-uniform axial field ($B_{p} \neq$ constant) can be super-Alfv\'enic
(even super-fast magnetosonic) in the downstream. We thus suggest
\beqn
V_{z} \propto z^{2/a}.
\eeqn
Again, $a$ denotes the power-law index of the parabolic streamline $z
\propto r^{a} (1 < a \le 2)$ of a magnetic flux tube. In the specific
case of a purely parabolic stream with $a=2$, we note the
trans-Alfv\'enic jet may exhibit that the flow speed increases linearly
as a function of the distance $V_{z} \propto z$, which can be observed
in NRMHD simulations \citep[{\em e.g.},][]{C95, OP97}.

In Fig. \ref{fig: bulk}, we show the bulk speed $\beta(=V/c)$ corrected
by Eq. (\ref{eq: V_app-V_int}) (again, adopting $\theta_{\rm
v}=14^{\circ}$) under an assumption of Eq. (\ref{eq: V_int-V}) as a
function of the de-projected distance, showing a feature of the
acceleration from $\beta \simeq 10^{-2}$ to $0.5$ in the range of
$z=10^{2}$ -- $10^4\,r_{\rm s}$. A distribution of $\beta$ exhibits a
linear increase as a function of $z$ as $\beta \propto z$ around $z
\simeq 10^{2}$ -- $10^3\,r_{\rm s}$, suggesting a trans-Alfv\'enic
flow. In order to fit the data points and derive a solid picture, more
sampling from this region is desired.  Therefore, higher angular
resolution and higher sensitivity observation with detailed analysis to
identify more components and trace the proper motions is
important. Beyond $z \gtrsim 10^{3}\,r_{\rm s}$, it seems that the bulk
acceleration is weakened slightly, indicating that the jet
may transit from trans-Alfv\'enic to trans-fast magnetosonic, but the
underlying flow is still accelerated toward a relativistic regime $\beta
\lesssim 1$.

\subsection{MHD Nozzle Equation in the Trans-magnetosonic Flow}
\label{sec: MHD-Nozzle}

Of our particular interest is the trans-(fast) magnetosonic flow in the
parabolic streamline. While the fraction of magnetic energy converted
into bulk kinetic energy may still be small at the CFP, a further
conversion from magnetic to kinetic energy and therefore a continued
bulk acceleration would take place beyond the CFP via the so-called
the ``magnetic nozzle'' effect, which has originally been pointed out by
\citet[]{C89}; poloidal magnetic flux is required to diverge
sufficiently rapidly in order that most of the Poynting flux in the
initially magnetically dominated outflow can be converted into the
kinetic energy flux beyond the CFP. This fundamentally important
hypothesis has been confirmed in steady, axisymmetric solutions in both
RMHD \citep[]{LCB92, L93, V04} and NRMHD \citep[]{KS97} regimes.

We examine a steady ($\partial/\partial t = 0$) and axisymmetric
($\partial/\partial \phi = 0$), NRMHD outflow along the poloidal
streamline in cylindrical coordinate ($z, \phi, z$). The system of ideal
MHD equations consists of the continuity, momentum (including the gravity),
entropy, and Maxwell equations and Ohm's law
\citep[{\em e.g.},][]{L87}:
\beqn
\label{eq: mass}
&&\nabla \cdot (\rho \bolV)=0, \\
\label{eq: momentum}
&&\rho(\bolV \cdot \nabla)\bolV=-\nabla p + \frac{1}{c}\bolJ \times \bolB - \rho \nabla \phi_{\rm g}, \\
\label{eq: entropy}
&&\bolV \cdot \nabla \left(p/\rho^{\Gamma}\right)=0, \\
\label{eq: Maxwell}
&&\nabla \times \bolB=\frac{4 \pi}{c}\bolJ, \ \nabla \times \bolE = 0, \ \nabla \cdot \bolB = 0, \\
\label{eq: Ohm}
&&\bolE +\frac{\bolV}{c} \times \bolB=0,
\eeqn
where $\Gamma$, $\phi_{\rm g}=-GM_{\bullet}/(r^{2}+z^{2})^{1/2}$,
$\bolJ$, and $\bolE$ are the specific heat ratio, gravitational
potential, current density, and electric field, respectively.  By using
eq.\,(\ref{eq: mass}) and the solenoidal constraint in eq.\,(\ref{eq:
Maxwell}), the velocity field can be split into poloidal and toroidal 
components as
\beqn 
\bolV = \bolV_{p}+V_{\phi}
\hat{\bole}_{\phi}, 
\eeqn 
with 
\beqn 
\label{eq: V} 
\bolV_{p}&=&V_{r} \hat{\bole}_{r} + V_{z} \hat{\bole}_{z} \nonumber \\
&=&\frac{1}{4 \pi \rho r}\left(-\frac{\partial \Phi}{\partial z} \hat{\bole}_{r} +
\frac{\partial \Phi}{\partial r} \hat{\bole}_{z}\right), 
\eeqn
where $\Phi (r, z)$ is the Stokes' poloidal stream function and constant
along a streamline.

\begin{figure}
\begin{center}
\includegraphics[scale=0.35, angle=-0]{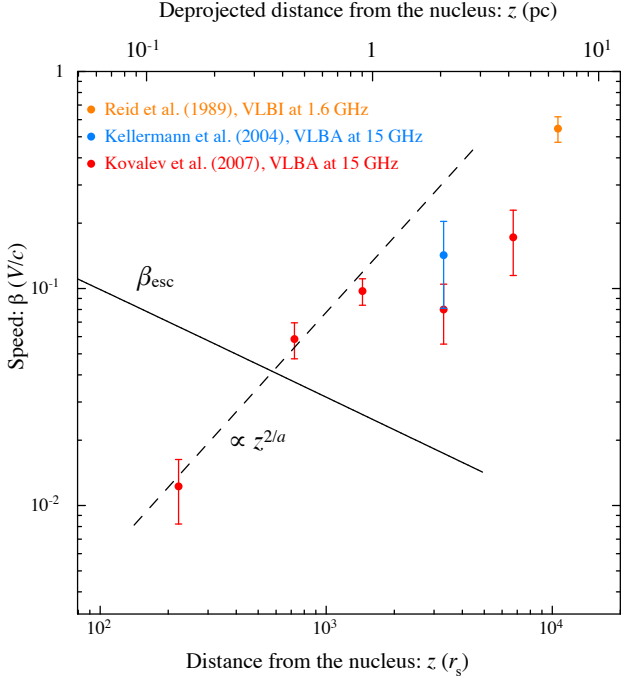} \caption{
\label{fig: bulk} 
Distribution of the bulk speed $\beta (=V/c)$ as a function of the
de-projected distance from the core. Measurements are illustrated in the
same manner as Fig. \ref{fig: proper}. The dashed line represents $\beta
\propto z^{2/a}$ with $a=1.73$; note it is not a fit to data points, but
is derived analytically in \S \ref{sec: TAW}. The solid line shows the
escape speed $\beta_{\rm esc} (=V_{\rm esc}/c) = 1/\sqrt{z}$ (normalized
by $r_{\rm s}$) along the jet parabolic (solid) streamline shown in
Fig. \ref{fig: r-z}.
}
\end{center}
\end{figure}

We consider that the super-Alfv\'enic jet carries a substantial energy
in the form of Poynting flux along the axial ($z$) direction, especially
for a limit of $B_{\phi} \gg B_{p}$ \citep[{\em in other words}, the
system can be described as a current-carrying jet, {\em e.g.},][]{C95}.
As is mentioned above, the divergence of poloidal magnetic flux is one
of key factors to attain an efficient conversion of the Poynting flux
into the kinetic energy flux. \citet[]{KS97} investigate two cross
sections of a poloidal magnetic flux $\Sigma$ in steady, axisymmetric
jets; the solution for $\Sigma \propto r^{2}$ almost fails to convert
the Poynting flux into the kinetic energy flux beyond the CFP so that no
further bulk acceleration occurs, remaining the Poynting-flux dominated
(two-thirds of the total energy) at $z/r_{0} \sim 10^{6}$, where $r_{0}$
denotes the radius at the foot point of the jet. On the other hand, if
the poloidal magnetic flux expands more than the purely parabolic case
beyond the Alfv\'en radius ($\Sigma \propto r^{2}+r^{3}/r_{\rm A}$), the
bulk acceleration continues until the Poynting flux decreases to zero,
well beyond the CFP at around $z/r_{0} \simeq 10^{5}$.

$r$-dependence of $\Sigma$ defines a divergence of the poloidal magnetic
flux, which is physically related to the degree of the magnetic
collimation of poloidal magnetic flux by the hoop
stress. Three-dimensional NRMHD simulations of current-carrying jets
exhibit the distribution of axial component of the magnetic field
becomes steeper, $B_{z} \propto r^{-3}$ in super-Alfv\'enic regime
\citep[]{NM04}. In general, the inwardly directed hoop stress (magnetic
tension) becomes dominant in the Lorentz forces at the inner core
(spine), while the outwardly directed magnetic pressure, which is
dominant in the Lorentz forces at the outer edge (sheath), will be
balanced with the external pressure so that the boundary shape of jets
may not be magnetically self-determined (the hoop stress plays a minor
role), but be controlled by the ambient ISM \citep[]{NLL06} as we will
also discuss in \S \ref{sec: Dis-parabolic}. In late years, RMHD
simulations confirm the intrinsic correlation between an efficiency of
bulk accelerations and non-uniform lateral expansions of poloidal
magnetic flux; it is referred as ``differential bunching'' of field
lines \citep[]{T09} or ``differential collimation'' \citep[]{K11}, but
they are essentially same as the magnetic nozzle effect \citep[]{C89}.

\begin{figure}
\begin{center}
\includegraphics[scale=0.5, angle=0]{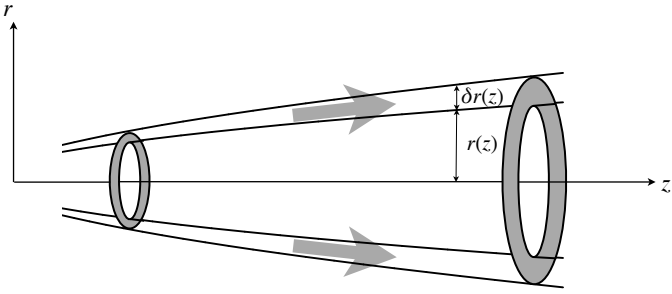} 
\caption{\label{fig: nozzle}
Schematic picture of our MHD nozzle model. The nozzle consists of a
hollow parabolic tube, but its thickness $\delta r (z)$ increases as a
function of the axial distance $z$, based on our consideration that the
flow is trans-Alfv\'enic to trans-fast magnetosonic; the hoop stress
differentially bunches the poloidal magnetic flux toward the central
axis by the sweeping magnetic twist magnetic twist process
\citep[]{US86}.}
\end{center}
\end{figure}

An edge-brightened feature is widely confirmed in the M87 jet from VLBI
to VLA scales \citep[]{R89, O89, J99, KOV07, L07, H11, AN12}, indicating
a hollow structure.  Schematic picture of our MHD nozzle model is shown
in Fig. \ref{fig: nozzle}. The jet expands in a parabolic shape $z
\propto r^{a} (1 < a \leq 2)$ with a dominant axial component of the
velocity field $V_{z}$.  Our domain of examination is a hollow tube with
a variable thickness $\delta r (z)$, which increases as a function of
$z$. We assume that the unit normal to the area surface is parallel to
the $z$-axis. A concept of the differential collimation means a gradual
concentration of poloidal magnetic flux towards the central part of
the flow; a faster collimation of the inner flux surfaces than of the
outer ones. This is mathematically expressed \citep[{\em e.g.},][]{K11}
as
\beqn
\frac{d}{d z} \left(\frac{\delta r}{r}\right) > 0,
\eeqn
which is equivalent to
\beqn
\label{eq: expansion}
\frac{1}{\delta r}\frac{d \delta r}{d z} >
\frac{1}{r}\frac{d r}{d z}.
\eeqn

Following \citet[]{C95} \citep[{\em see also},][]{LS97},
we apply Gauss' divergence theorem to eq.\,(\ref{eq: mass})
with a cross-sectional area $r \delta r$:
\beqn
\label{eq: mass-flux}
\rho V_{z} r \delta r = {\rm constant} = \kappa(\Phi).
\eeqn
Faraday's law in Eq.\,(\ref{eq: Maxwell}) together with Eq.\,(\ref{eq: Ohm})
implies the ``frozen-in'' field equation $\nabla \times (\bolV \times \bolB)=0$.
Using Stokes' theorem, we derive
\beqn
\label{eq: frozen-in}
V_{z} B_{\phi} \delta r = {\rm constant} = \chi(\Phi).
\eeqn
The entropy equation\,(\ref{eq: entropy}) gives the adiabat 
\beqn
\label{eq: adiabat}
p/\rho^{\Gamma} = {\rm constant} \equiv Q (\Phi),
\eeqn
where above functions $\kappa$, $\chi$, and $Q$ of $\Phi$ are 
conserved quantities along a flow streamline.

Considering eq.\,(\ref{eq: momentum}) in the one-dimensional axial ($z$)
direction at $z \gg r$, we can use the following equation:
\beqn
\label{eq: momentum-z}
\rho V_{z} \frac{d V_{z}}{d z}=-\frac{d p}{d z}-\frac{d}{d z}
\left(\frac{B_{\phi}^2}{8 \pi}\right)-\rho \frac{G M_{\bullet}}{z^2}.
\eeqn
By differentiating eqs.\,(\ref{eq: mass-flux}) - (\ref{eq: adiabat}) and
substituting them into eq.\,(\ref{eq: momentum-z}), we obtain 
\beqn
\label{eq: Hugoniot-pre1}
(V_{z}^2-V_{f}^2)\frac{1}{V_z}\frac{d V_z}{d z}=\frac{V_{f}^2}{\delta
r}\frac{d \delta r}{d z}+\frac{C_{s}^2}{r}\frac{d r}{d z}-\frac{G M_{\bullet}}{z^2},
\eeqn 
where $V_{f}^2=C_{s}^2+V_{\rm A \phi}^2$ denotes the classical fast magnetosonic
speed ($C_{s} \equiv \Gamma p/\rho$ and $V_{\rm A \phi} \equiv
B_{\phi}^{2}/4 \pi \rho$ are the sonic and toroidal Alfv\'en
speeds). Note that we eliminate the second term in the right hand side
of eq.\,(\ref{eq: Hugoniot-pre1}) because of eq.\,(\ref{eq: expansion})
and our assumption, $V_{f} \simeq V_{\rm A \phi}$ in the flow:
\beqn
\label{eq: Hugoniot-pre2}
(V_{z}^2-V_{f}^2)\frac{1}{V_z}\frac{d V_z}{d z}=\frac{V_{f}^2}{\delta
r}\frac{d \delta r}{d z}-\frac{G M_{\bullet}}{z^2}.
\eeqn 

Now supposing that $z \propto \delta r^{a} (1 < a \leq 2)$, then our
approximate Hugoniot equation for the parabolic MHD nozzle can be
described as
\beqn
\label{eq: Hugoniot}
\frac{z}{V_{z}}\frac{d V_{z}}{dz} \approx 
\frac{V_{f}^2/a-G M_{\bullet}/z}{V_{z}^2-V_{f}^2}.
\eeqn
Thus, for a specific case $a=2$, we speculate the poloidal jet speed
exceeds the local escape speed $V_{\rm esc}$ around trans-fast
magnetosonic regime: \beqn V_{z} \approx V_{f} \approx V_{\rm esc},
\eeqn where $V_{\rm esc}=\sqrt{2GM_{\bullet}/z}$. As is discussed in
\citet[]{KS97}, when a cross section expands more rapidly than $\Sigma
\propto r^2$, the classical fast magnetosonic point approaches near the
Alfv\'en point by an enhancement of bulk accelerations in the diverging
magnetic nozzle effect, but the fast magnetosonic point is fairly bound
to around to the location around $1/2 V_{p}^2 \simeq \phi_{ \rm g}$ in
both cases ($\Sigma \propto r^{2}$ and $\Sigma_{\infty} \propto r^3$).

Note eq. (\ref{eq: Hugoniot}) represents an identical
formula (when $a=2$) with the equation of the hydrodynamic solar wind
\citep[]{P58} with replacing $C_{s}$ with $V_{f}$ (the thermal pressure
is substituted by the toroidal magnetic pressure). Thus, the nature of
the trans-fast magnetosonic flow in a parabolic MHD nozzle, which is
powered by TAWs, is similar to the one of transonic solution of Parker's
solar wind. The hoop stress differentially bunches the poloidal magnetic
flux toward the central axis so that a rapid expansion of the cross
sectional area of the poloidal magnetic flux operates on the hollow
parabolic MHD stream; we speculate a de Laval-like \citep[]{D67} 
diverging nozzle effect in the super-fast magnetosonic regime (see, also
Fig. \ref{fig: nozzle}).

The distribution of $V_{\rm esc}$ along the parabolic streamline of the
M87 jet is drawn in Fig. \ref{fig: bulk}, showing the location where the
jet is trans-escape as well as trans-fast magnetosonic. It is notable
that the jet obtains a speed of $\sim 0.04\,c$ in order to escape under
control of the gravity of the SMBH at around 0.2 -- 0.3 pc ($\simeq 6
\times 10^2\,r_{\rm s}$). Thus, we speculate, as a consequence of the
energy conversion from Poynting to kinetic energy fluxes, the MHD bulk
acceleration takes place beyond a super-fast magnetosonic regime; the
magnetic nozzle effect \citep[]{C89} is being activate by the sweeping
magnetic twist mechanism \citep[]{US86} so that a further energy
conversion from Poynting to kinetic energy fluxes can be expected in the
parabolic MHD nozzle as examined above.

If the jet is highly magnetized ($V_{\rm A \phi}^2 \gg
C_{s}^2$) and/or a cross section of the poloidal magnetic flux expands
more rapidly than $\Sigma \propto r^2$, then a position of the classical
fast magnetosonic point approaches near the Alfv\'en point as mentioned
above. This could be the case in the M87 jet. Thus, we may not
distinguish clearly the trans-Alfv\'enic region from the
trans-magnetosonic region in Fig. \ref{fig: bulk}. However, based on an
intrinsic nature of MHD outflows, the accelerated flow first becomes
trans-Alfv\'enic, and then trans-magnetosonic in the dynamical
evolution. Note that the distribution of the bulk speed in
Fig. \ref{fig: bulk} may indicate a transition from trans-Alfv\'enic to
trans-magnetosonic regime at around $\sim 10^{3}\,r_{\rm s}$, where the
speed becomes comparable to the local escape speed. A sequential MHD
bulk acceleration, which can be also powered by the toroidal magnetic
pressure gradient, maintains even after weakening of the acceleration at
super-Alfv\'enic regime as we examine the magnetic nozzle effect above.

\section{Discussions}
\label{sec: Dis}

\subsection{Formation of the Parabolic Streamline}
\label{sec: Dis-parabolic}
It is widely believed that AGN jets are highly supersonic from their
earlier evolution stage nearby the central engine with being a narrowly
beamed conical structure. Because the jet interior is out of causal
contact with its outer edges by sonic waves in the lateral direction, a
freely expanding ballistic flow is considered. We, however, suggested
that the M87 jet is not a case of the conical jet, but the parabolic jet
up to over $10^{5} \, r_{\rm s}$ (AN12). Furthermore, if we
consider MHD jets in general, which is one of popular theoretical
framework to organize the jet dynamics, it may be the norm that AGN jets
are still in causal contact in the lateral expansion through a
communication by magnetosonic waves.

MHD jets can be a parabolic $z \propto r^{a} (1 < a \le 2)$ if the ISM
pressure is decreasing as $p_{\rm ism} \propto z^{-b}$ with $b \simeq 2$
\citep[{\em e.g.}\rm,][]{T08, Z08, K09, L10}. On the other hand, the jet
boundary does {\em not} adjust to the ambient ISM pressure profile,
instead asymptote to a conical shape for $b > 2$ \citep[{\em
e.g.}\rm,][]{Z08, K09, L10, KB12}. This indicates that the jet lateral
expansion time becomes shorter than the signal propagation time of
magnetosonic waves across the bulk flow, leading to a loss of causal
contact to form a freely expanding ballistic jet. Non-conical, parabolic
streamlines require a transverse pressure support by the extremal ISM
\citep[]{BR74, BK79}. Thus, we suggest that the M87 jet may be {\em
magnetically accelerated, but thermally confined by the ISM} (rather
than the self-confined in the MHD jet dynamics) as we will discuss
below.

A power-law like distribution of the ISM pressure with $b \sim 2$ would
be expected towards the ISM environment of M87 inside the Bondi
accretion radius $R \lesssim R_{\rm B}$ ($R$ denote the in a spherical
geometry) as a consequence of hot RIAFs, which are generally considered
in low luminosity AGNs. During past several decades, various
type of RIAF models has been examined and we can now categorized them
into three main models; the advection-dominated accretion flow
\citep[ADAF, {\em e.g.},][]{I77, NY94}, the convection-dominated
accretion flow \citep[CDAF, {\em e.g.},][]{NIA00, QG00}, and the
advection-dominated inflow-outflow solution \citep[ADIOS, {\em
e.g.},][]{BB99, BB04}.

Given the self-similar analytical treatment among these RIAF models, the
gas density distribution can be described as a power-law scaling
$\rho_{\rm ism} \propto R^{-k}$, where $k=3/2$ for \citet[]{B52}
accretion flow (BAF) and ADAF, in which the gas undergoes a nearly
free-fall. On the other hand, $k=1/2$ in CDAF is found because the
radial convection transports the angular momentum inwards, causing a
reduction of the mass accretion rate and a much flatter radial density
profile. ADIOS model is known to have intermediate values $1/2 < k <
3/2$ by taking into account a wind carrying away the mass, angular
momentum, and energy. Assuming the polytropic equation of state, $p_{\rm
ism} \propto z^{-b}=R^{-(k+1)}$ (we here denote $z=R$). Thus, $1.5 \le b
\le 2.5$ are allowed in RIAF models.

A recent analytical model for a giant ADAF \citep[]{NF11} shows $b \sim
2.3$ from beyond $R_{\rm B} \simeq 5 \times 10^{5} \, r_{\rm g}$ down to
the SMBH, and it is universally derived for slow rotation of the ISM
($\lesssim 30 \, \%$ of the Keplerian speed) with the wider range of the
$\alpha$-viscosity parameter \citep[0.001 - 0.3,][]{SS73}. A spherically
symmetric accretion with MHD turbulence \citep[]{S08} produces $b \sim
2.25$ without a strong dependence on the parameters and the equation of
state. On the other hand, thanks to comprehensive studies of numerical
simulations over last decade by many authors for exploring nonlinear
dynamics on RIAFs, we learn that the density profile flattens compared
to the original BAF/ADAF analytical solutions, but the system does not
approach a full CDAF, staying at the intermediate ADIOS, $\rho_{\rm ism}
\propto R^{-k}$ with $k \lesssim 1$, corresponding to $p_{\rm ism}
\propto z^{-b}$ with $b \lesssim 2$ \citep[{\em e.g.},][for recent
progresses on numerical simulations of RIAFs and useful references
therein]{Y12a, Y12b}.

Spatial resolutions towards M87 in current X-ray instruments may not be
adequate enough to resolve the region ($\lesssim 1\arcsec$) inside the
Bondi accretion radius of $3.8 \times 10^{5} \, r_{\rm s}$ ($\sim 250$
pc) although {\em Chandra} X-ray observations confirmed ISM properties,
such as the King core radius $r_{\rm c} \simeq 1.4$ kpc and the Bondi
radius $R_{\rm B} \sim$ 250 pc \citep[see, {\em e.g.},][]{YWM02,
D03}. We, however, remark that {\em Chandra} X-ray observations revealed
the RIAF-like feature in a nearby low luminosity AGN, NGC 3115
\citep[]{W11}. Stellar kinematics of the bulge of NGC 3115 give the SMBH
mass estimation as $M_{\bullet} \sim (1-2) \times 10^{9} M_{\odot}$,
identifying this source to be the nearest $> 10^{9} M_{\odot}$ SMBH
because of its proximity $D \sim 9.7$ Mpc \citep[][and references
therein]{W11}.  The Bondi radius ($R_{\rm B} \sim 4\arcsec - 5\arcsec$)
is spatially resolved for the first time in AGNs; the radial density
profile within $4\arcsec$ is described as $\rho_{\rm ism} \propto
R^{-1.03^{+0.23}_{-0.21}}$, giving a nice agreement with the RIAF model.
Thus, we speculate that $p_{\rm ism} \propto z^{-b}$ with $b \sim 2$ may
be moderately reasonable.

\subsection{A Nature of the upstream flow within 100
  $r_{\rm s}$}
\label{sec: Dis-upstream}

We first discuss about the hot accretion disk as an origin of the MHD
outflow in M87. In ADAF (RIAF) model ($C_{s} \sim V_{\rm K}$, where the
Keplerian speed $V_{\rm K}$ is equal to $c/\sqrt{2 r/r_{\rm s}}$), the
disk (ion) temperature is considered as nearly virial $T_{\rm i} = 5.4
\times 10^{12} (r/r_{\rm s})^{-1}$ K \citep[]{NY94}. Given $r \sim 10\,
r_{\rm s}$ in Fig. \ref{fig: r-z} (where VLBI cores are located), the
sound speed $C_{s} \simeq 0.28\, c$ is also comparable to the local
escape speed $V_{\rm esc} = c/\sqrt{r/r_{\rm s}} \sim 0.3\,
c$. Therefore, we would seek much faster motions in the downstream than
measured speeds $0.01\,c$ -- $0.1\,c$ at $10^{2} - 10^{3} \, r_{\rm s}$
based on over a ten-year VLBA observations at 15 GHz \citep[][{\em see
also} Fig. \ref{fig: bulk}]{KOV07}, which are considered as the bulk
speed in \S \ref{sec: Proper}.  If the M87 jet is indeed originated from
the surface of RIAFs \citep[$B_{\phi} \sim B_{p}$; {\em e.g.},][]{M01},
as referring \citet[]{PMM13} in \S \ref{sec: MHD-Jets}, we may not
expect a standard magneto-centrifugal acceleration process (BP82) in the
sub-Alfv\'enic region, but favor an initially trans-Aflv\'enic flow
\citep[{\em see also},][]{C95}. Thus, this may give some constraint on
the required MHD jet model with an initial jet speed with an order of
$\sim 0.1\, c$\footnote{\citet[]{L07} reported possible, but
transient features $V_{\rm app}=$ (0.25--0.40) $c$ located at about 3
mas based on just two of five total epochs ($\sim$ 8 months apart) of
VLBA and global VLBI observations at 43 GHz; these components would not
be present in other epochs. It may be still unclear whether a detection
of an order of $V=0.5\, c - 0.6\, c$ (at $\simeq 1.5\times 10^3\, r_{\rm
s}$ in de-projection) will support a RIAF-originated jet scenario.} at
the jet footpoint.

Next, we discuss about the Kerr black hole as an origin of the MHD
outflow in M87. As we introduced in \S \ref{sec: MHD-Jets}, the
outermost sheath of the ergosphere-driven jet exhibits non-relativistic
motion $\gamma \simeq 1$ at $\sim 5 \times 10^3\,r_{\rm g}$ as seen
GRMHD simulations \citep[]{M06}; from a polar angle cross-sectional
distribution of the gas density, we find that the outermost sheath is
much higher mass-loading (about four orders of magnitude) than the
innermost jet spine. This may cause an inefficient bulk acceleration at
least up to this spatial scale. A matter piling up at the outer surface
of the jet may also be responsible for observed edge-brightened features
in M87. The location of the CFP is about a few of $100\, r_{\rm s}$ near
the surface layer of the ergosphere-driven jet where the jet is still
Poynting-flux dominated. Thus, we may speculate a behavior of the jet at
sub- to trans-Alfv\'enic region $z \lesssim 100\, r_{\rm s}$ as a
magneto-centrifugal driving process (by not a disk rotation, but an
ergosphere) similar to BP82 and then the jet seems to be powered by TAWs
(the toroidal magnetic pressure gradient force) at trans-Alfv\'enic to
trans-fast magnetosonic regions. Note that a magneto-centrifugal driving
allows the flow to start at a very slow speed $V_{0}$ compared to the
rotational motion of the magnetic field \citep[]{C95}. In the case of
non-relativistic Keplerian disk, $V_{\rm K}/V_{0} \simeq 300$ can be
obtained in steady, axisymmetric NRMHD outflow solutions \citep[]{KS97,
VTST00}. Thus, even relativistic field rotations with $0.1\, c$ -- $c$
nearby the black hole, we may speculate an initial jet speed with $V_{0}
\ll 0.1\, c$ at the jet footpoint.

One of crucial tests to constrain a nature of the upstream flow within
100 $r_{\rm s}$ is to image extended jet emissions from the VLBI core.
We are conducting new GMVA observations at 86 GHz in order to resolve
innermost jet structure at a few of $10^{1-2}\, r_{\rm s}$ scales.

\subsection{An Offset of the Jet Foot Point from the Black Hole and 
Effect on Frequency Depending Core Shift Measurements}
\label{sec: Dis-offset}

In \S \ref{sec: Res}, we show three additional data points of VLBI cores
seem to fairly follow a parabolic streamline derived by AN12,
implying that the jet maintains a single non-conical structure with
nearly five orders of magnitude in distance. We here discuss about
limitations of a frequency depending VLBI core shift measurement to seek the
true origin of the M87 jet. The median values of the VLBI core shifts in AGN
jets are reported as 0.080 mas \citep[]{S11} and 0.128 \citep[]{P12}
between 15.4 -- 8.1 GHz. A scale of $\sim 0.1$ mas corresponds to 1-10 pc
in de-projection for typical blazars with a small viewing angle
$\theta_{\rm v} \sim 5^{\circ}$, which is equivalent to
$10^{3-4}\,r_{\rm s}$ with $M_{\bullet}=10^{9} M_{\odot}$. Therefore, if one
applies this method to blazars at even higher frequencies, an axial
offset between the position of the SMBH (the origin) and/or accretion
disk (the equatorial plane) and the jet foot point may be ignorable.

On the other hand, in the case of M87, the value of the VLBI core shift
between 15.4 -- 8.4 GHz is similar \citep[$\sim 0.1$ mas;][]{H11}, but it
corresponds to $\sim 10\,r_{\rm s}$, approaching a limit to apply the method
of frequency depending VLBI core shift measurements. By using eq. (\ref{eq:
core-shift}), we derive $\Delta z$ (230 GHz) = $4.3 \pm 2.2\,r_{\rm s}$, which may
be comparable to the so-called stagnation surface (SS) seen in general
relativistic MHD simulations \citep[{\em e.g.},][]{M06}. If we consider
the MHD jet launching from the rotating SMBH magnetosphere
\citep[]{BZ77}, the SS (where the poloidal velocity component changes
the sign $-/+$; the inflow onto the SMBH inside the SS, while the
outflow as a jet outside the SS) has an offset from the SMBH. The
location of the SS depends on a spin of the black hole; the balance
between the gravity and the centrifugal forces along a poloidal magnetic
field in the co-rotating frame with the line of force.

\citet[]{M06} prescribes a black hole spin of $j=0.9375$ and the SS is
located at around $2-3\,r_{\rm s}$. A lower spin will put the radius of the
SS further downstream $> \Delta z$ (230 GHz). Even if we consider that
the MHD jet originates from a RIAF-type accretion disk $H \sim r,
r\gtrsim 2.5\,r_{\rm s}$ \citep[the lensed ISCO radius for the prograde
spinning $j=1$;][]{D12}, then the disk surface is presumably the SS so
that the situation will be unchanged. Thus it may be challenging to
explore the jet foot point in M87 by using VLBI cores at higher
frequencies. If we take an offset of the SS into consideration, three
VLBI cores may not be located on a parabolic streamline in
Fig. \ref{fig: r-z}. Thus, future sub-mm VLBI imaging of the jet foot
point together with the SMBH shadow is crucial.

\section{Summary}
\label{sec: Sum} Following \citet[]{AN12}, we further investigate the
parabolic jet structure in M87 with additional data points of VLBI cores
and examine theoretical aspects of MHD jets.  We here propose a
hypothesis that the observed subluminal motions on 0.1 -- 10 pc scales
in projection (corresponding to $10^{2}$ -- $10^{4}\,r_{\rm s}$) show
the MHD bulk acceleration in a non-relativistic regime although they
have been interpreted during past decades as standing shocks and/or some
patterns of plasma instabilities in supersonic jets. Based on our
analysis, it is reasonably described as a consequence of Poynting
flux-dominated flows, powered by large amplitude, nonlinear torsional
Alfv\'en waves in a parabolic streamline. We examine the
non-relativistic MHD nozzle equation in a parabolic shape. The nature of
trans-fast magnetosonic flow is similar to the one of transonic solution
of Parker's hydrodynamic solar wind; the jet becomes super-escape as
well as super-fast magnetosonic at around $\sim 10^{3}\,r_{\rm s}$,
while the upstream trans-Alfvenic flow speed increases linearly as a
function of the distance at $\sim 10^{2}$ -- $10^{3}\, r_{\rm s}$. We
here point out that this is the first evidence to identify these
features in astrophysical jets.

We discuss about a formation of the parabolic streamline as a
consequence of the lateral force balance between the jet internal
(magnetic) and the external ISM (thermal) pressures inside the sphere of
gravitational influence of the SMBH potential. An origin of the M87 jet
is also discussed based on our current understanding in observations and
theories although we will need further observational constraints of a
nature of the upstream flow within 100 $r_{\rm s}$. We furthermore
suggest that it may be challenging for a seeking the true jet origin
only by using VLBI core shift measurements in M87. Thus, we encourage
future observations with a high sensitivity such as a space VLBI and/or
mm/sub-mm VLBI at 86 GHz or higher. In the forthcoming paper, we will
examine a continued bulk acceleration from non-relativistic to fully
relativistic regime, which may occur at around $10$ pc ($\sim
10^{4}\,r_{\rm s}$) or further downstream toward the HST-1 complex so
that the we can complete the analysis of the parabolic jet stream line
in M87 in $10^{2}$ -- $10^{5}\,r_{\rm s}$ in a framework of MHD jets.

We appreciate an anonymous referee for critical, but very helpful
comments for revising the manuscript. We acknowledge A. Doi, M. Kino,
H. Nagai, and K. Hada, for useful comments and suggestions. MN thanks
D.~L. Meier and C.~A. Norman for stimulating discussions on the MHD jet
theory and Z.-Y. Li for giving useful comments on the magnetic nozzle
effect in the manuscript. We are grateful to M. Inoue, P.~T.~P. Ho,
S. Matsushita, and our members on the Greenland Telescope project of the
Academia Sinica, Institute of Astronomy and Astrophysics for their warm
encouragements.

%{\it Facilities:} \facility{VLBA, EHT}.

\begin{table*}
\begin{center}
\caption{Qualities of the images}
  \begin{tabular}{@{}lcccccc@{}}
  \hline
  \hline
   Frequency & & Synthesised Beam  & & Peak Intensity & rms Noise Level & Dynamic Range \\

  (GHz) & (mas) & (mas) & (degree) & (Jy beam$^{-1}$) & (mJy
                       beam$^{-1}$) & \\
  \hline
  43 & 0.281 & 0.139 & -0.07$^{\circ}$ & 0.463 & 0.63 & 735 \\
  86 & 0.406 & 0.212  & 81.6$^{\circ}$ & 0.884 & 8.73 & 101 \\
  \hline
\end{tabular}
\label{table:QL}
\end{center}
\end{table*}


\begin{thebibliography}{}
\bibitem[Achterberg \etal (1983)]{ABG83}
Achterberg, B., Blandford, R.~D., \& Goldreich, P. 1983, \nat, 304, 607
\bibitem[Asada \& Nakamura (2012)]{AN12}
Asada, K., \& Nakamura, M., 2012, \apj, 745, L28 (AN12)
\bibitem[Biretta et al. (1995)]{B95}
Biretta, J.~A., Zhou, F., \& Owen, F.~N. 1995, \apj, 447, 582
\bibitem[Biretta \etal (1999)]{B99}
Biretta, J.~A., Sparks, W.~B., \& Macchetto, F. 1999, \apj, 520, 621
\bibitem[Blandford \& Begelman (1999)]{BB99}
Blandford, R.~D., \& Begelman M.~C. (1999), \mnras, 303, L1
\bibitem[Blandford \& Begelman (2004)]{BB04}
Blandford, R.~D., \& Begelman M.~C. (2004), \mnras, 349, 68
\bibitem[Blandford \& K\"{o}nigl (1979)]{BK79}
Blandford, R.~D., \& K\"{o}nigl, A. 1979, \apj, 232, 34
\bibitem[Blandford \& Payne (1982)]{BP82}
Blandford, R.~D., \& Payne, D.~G. 1982, \mnras, 199, 883 (BP82)
\bibitem[Blandford \& Rees (1974)]{BR74}
Blandford, R.~D., \& Rees, M.~J. 1974, \mnras, 169, 39
\bibitem[Blandford \& Znajek(1977)]{BZ77}
Blandford, R.~D., \& Znajek, R.~L. 1977, \mnras, 179, 433
\bibitem[Bondi(1952)]{B52}
Bondi, H. 1952, \mnras, 112, 195
\bibitem[Camenzind (1989)]{C89}
Camenzind, M. 1989, in {\em Accretion Disks and Magnetic Fields in
                Astrophysics}, ed. G. Belvedere (Cordrecht: Kluwer), 129
\bibitem[Cheung et al. (2007)]{C07}
Cheung, C.~C., Harris, D.~E., \& Stawarz, \L. 2007, \apj, 663, L65
\bibitem[Contopoulos (1995)]{C95}
Contopoulos, J. 1995, \apj, 450, 617
\bibitem[Daly \& Marscher (1988)]{DM88}
Daly, R.~A., \& Marscher, A.~P. 1988, \apj, 334, 539
\bibitem[Dessler (1967)]{D67}
Dessler, A.~J. 1967, Rev. Geophys. 5, 1
\bibitem[Di Matteo et al. (2003)]{D03}
de Matteo, T., Allen, S.~W., Fabian, A.~C., Wilson, A.~S., 
\& Young, A.~J. 2003, \apj, 582, 133
\bibitem[Doeleman \etal (2012)]{D12}
Doeleman, S.~S., \etal 2012, Science, 338, 355
\bibitem[Gebhardt \etal (2011)]{G11} Gebhardt, K., Adams,
J., Richstone, D., \etal 2011, \apj, 729, 119
\bibitem[Gracia \etal (2009)]{G09}
Gracia, J., Vlahakis, N., Agudo, I., Tsinganos, K., 
\& Bogovalov, S.~V. 2009, \apj, 695, 503
\bibitem[Hada \etal (2011)]{H11}
Hada, K., \etal 2011, \nat, 477, 185
\bibitem[Harris \etal (2006)]{H06}
Harris, D.~E., Cheung, C.~C., Biretta, J.~A., Sparks, W.~B., 
Junor, W., Perlman, E.~S., \& Wilson, A.~S. 2006, \apj, 640, 211
\bibitem[Heyvaerts \& Norman (1989)]{HN89}
Heyvaerts, J., \& Norman, C.~A. 1989, \apj, 347, 1055
\bibitem[Heyvaerts et al. (2012)]{HLM12}
Heybaerts, J., Lehner, T., \& Mottez, F. 2012, \aap, 542, A128
\bibitem[Ichimaru (1977)]{I77}
Ichimaru, S. 1977, \apj, 214, 840
\bibitem[Junor \etal (1999)]{J99}
Junor, W., Biretta, J.~A., \& Livio, M. 1999, \nat, 401, 891
\bibitem[Jord{\'a}n \etal (2005)]{J05} Jord{\'a}n, A.,
C{\^o}t{\'e}, P., Blakeslee, J.~P., \etal 2005, \apj, 634, 1002
\bibitem[Jorstad \etal (2007)]{J07}
Jorstad, S.~G., \etal 2007, \aj, 134, 799
\bibitem[Kellermann \etal (2004)]{K04}
Kellermann, K.~I., \etal 2004, \apj, 539, 563
\bibitem[Kigure \& Shibata (2005)]{KS05}
Kigure, H., \& Shibata, K. 2005, \apj, 634, 879
\bibitem[Kohler \& Begelman (2012)]{KB12}
Kohler, S., \& Begelman, M.~C. 2012, \mnras, 426, 595
\bibitem[Kormendy \etal (1996)]{K96}
Kormendy, J., \etal 1996, \apj, 459, L57
\bibitem[Komissarov \etal (2007)]{K07}
Komissarov, S.~S., Barkov, M.~V., Vlahakis, N., \& K\"{o}nigl, A.
2007, \mnras, 380, 51
\bibitem[Komissarov \etal (2009)]{K09}
Komissarov, S.~S., Vlahakis, N., K\"{o}nigl, A.,
Barkov, M.~V. 2009, \mnras, 394, 1182
\bibitem[Komissarov (2011)]{K11}
Komissarov, S.~S. 2011, Mem. Soc. Astron. Ital., 82, 95
\bibitem[K\"{o}nigl \& Pudritz (2000)]{KP00}
K\"{o}nigl, A. \& Pudritz, R.~E. 2000, Protostars and Planets IV,
ed. V. Mannings \etal (Tucson, AZ: Univ. Arizona Press), 759
\bibitem[Kovalev \etal (2007)]{KOV07}
Kovalev, Y.~Y., Lister, M.~L., Homan, D.~C., \& 
Kellermann, K.~I. 2007, \apj, 668, L27
\bibitem[K\"{o}nigl (1981)]{K81}
K\"{o}nigl, A. 1981, \apj, 243, 700
\bibitem[Krichbaum \etal (2006)]{K06}
Krichbaum, T.~P., \etal 2006, J. Phys. Conf. Ser., 54, 328
\bibitem[Kudoh \& Shibata (1997)]{KS97}
Kudoh, T., \& Shibata, K. 1997, \apj, 474, 362
\bibitem[Kudoh \etal (1998)]{KMS98}
Kudoh, T., Matsumoto, R., \& Shibata, K. 1998, \apj, 508, 186
\bibitem[Lamers \& Cassinelli (1999)]{LC99}
Lamers, H.~J.~G.~L.~M., \& Cassinelli, J.~P. 1999, 
{\em Introduction to Stellar Wind} (Cambridge: Cambridge Univ. Press)
\bibitem[Lee \etal (2008)]{L08}
Lee, S.-S., \etal 2008, \aj, 136, 159
\bibitem[Li \etal (1992)]{LCB92}
Li, Z.-Y., Chiueh, T., \& Begelman, M.~C., 1992, \apj, 394, 459
\bibitem[Li (1993)]{L93}
Li, Z.-Y. 1993, \apj, 415, 118
\bibitem[Liffman \& Siora (1997)]{LS97}
Liffman, K., \& Siora, A. 1997, \mnras, 290, 629
\bibitem[Lister \etal (2009)]{L09}
Lister, M., \etal 2009, \aj, 138, 1874
\bibitem[Lovelace \etal (1987)]{L87}
Lovelace, R.~V.~E., Wang, J.~C.~L., \& Sulkanen, M.~E. 1987, \apj, 315, 504
\bibitem[Lovelace \etal (1991)]{L91}
Lovelace, R.~V.~E., Berk, H.~L., \& Contopoulos, J. 1991, \apj, 379, 696
\bibitem[Ly \etal (2007)]{L07}
Ly, C., Walker, R.~C., \& Junor, W. 2007, \apj, 660, 200
\bibitem[Lyubarsky (2010)]{L10}
Lyubarsky, Y. 2010, \mnras, 402, 353
\bibitem[Macchetto \etal (1997)]{M97} Macchetto, F.,
Marconi, A., Axon, D.~J., \etal 1997, \apj, 489, 579
\bibitem[Marshall \etal (2002)]{M02}
Marshall, H.~L., Miller, B.~P., Davis, D.~S., Perlman, E.~S., 
Wise, M., Canizares, C.~R., \& Harris, D.~E. 2002, \apj, 564, 683
\bibitem[McKinney (2006)]{M06}
McKinney, J.~C. 2006, \mnras, 368, 1561
\bibitem[Meier (1999)]{M99}
Meier, D.~L. 1999, \apj, 522, 753
\bibitem[Meier \etal (2001)]{MKU01}
Meier, D.~L., Koide, S., \& Uchida, Y. 2001, Science, 291, 84
\bibitem[Meier (2001)]{M01}
Meier, D.~L. 2001, \apj, 548, L9
\bibitem[Nakamura \& Meier (2004)]{NM04}
Nakamura, M., \& Meier, D.~L. 2004, \apj, 617, 123
\bibitem[Nakamura \etal (2006)]{NLL06}
Nakamura, M., Li, H., \& Li, S. 2006, \apj, 652, 1059
\bibitem[Narayan \& Yi (1994)]{NY94}
Narayan, R., \& Yi, I. 1994, \apj, 428, 13L
\bibitem[Narayan \etal (2000)]{NIA00}
Narayan, R., Igumenshchev, I.~V., \& Abramowicz, M.~A. 2000, \apj, 539, 798
\bibitem[Narayan \& Fabian (2011)]{NF11}
Narayan, R., \& Fabian A.~C. 2011, \mnras, 415, 3721
\bibitem[Ouyed \& Pudritz (1997)]{OP97}
Ouyed, R., \& Pudritz, R.~E. 1997, \apj, 482, 712
\bibitem[Owen \etal (1989)]{O89}
Owen, F.~N., Hardee, P.~E., \& Cornwell, T.~J. 1989, \apj, 340, 698
\bibitem[Parker (1958)]{P58}
Parker, E.~N. 1958, \apj, 128, 664
\bibitem[Pelletier \& Pudritz (1992)]{PP92}
Pelletier, G., \& Pudritz, R.~E. 1992, 394, 117
\bibitem[Perlman \etal (1999)]{P99}
Perlman, E.~S., Biretta, J.~A., Zhou, F., Sparks, W.~B., \& 
Macchetto, F.~D. 1999, \aj, 117, 2185
\bibitem[Perlman \& Wilson (2005)]{PW05}
Perlman, E.~S., \& Wilson, A.~S. 2005, \apj, 627, 140
\bibitem[Polko \etal (2010)]{PMM10}
Polko, P., Meier, D.~L., \& Markoff, S. 2010, \apj, 723, 1343
\bibitem[Polko \etal (2013)]{PMM13}
Polko, P., Meier, D.~L., \& Markoff, S. 2013, \mnras, 428, 587
\bibitem[Priest (1981)]{P81}
Priest, E.~R. 1981, {\em Solar Magneto-hydrodynamics} (Peidel,
                Dordrecht), p. 160
\bibitem[Pudritz \& Norman (1983)]{PN83}
Pudritz, R.~E., \& Norman, C.~A. 1983, \apj, 274, 677
\bibitem[Pushkarev (2012)]{P12}
Pushkarev, A.~B., \etal 2012, \aap, 545, A113
\bibitem[Quataert \& Gruzinov (2000)]{QG00}
Quataert, E., \& Gruzinov, A. 2000, \apj, 545, 842
\bibitem[Reid \etal (1989)]{R89}
Reid, M.~J., Biretta, J.~A., Junor, W., Muxlow, T., \& Spencer, R. \apj,
                336, 112
\bibitem[Romanova \etal (1997)]{R97}
Romanova, M.~M., Ustyugova, G.~V.~U., Koldoba, A.~V., 
Chechetkin, V.~M.~C., \& Lovelace, R.~V.~E. 1997, ApJ, 482, 708
\bibitem[Sanders (1983)]{S83}
Sanders, R.~H. 1983, \apj, 266, 73
\bibitem[Shcherbakov (2008)]{S08}
Shcherbakov, R.~V. 2008, \apjs, 177, 493j
\bibitem[Shakura \& Sunyaev (1973)]{SS73}
Shakura, N.~I., \& Sunyaev R.~A, 1973, \araa, 24, 337
\bibitem[Shibata \& Uchida (1985)]{SU85}
Shibata, K., \& Uchida, Y. 1985, \pasj, 37, 31
\bibitem[Shibata \& Uchida (1986)]{SU86}
Shibata, K., \& Uchida, Y. 1986, \pasj, 38, 631
\bibitem[Sokolovsky \etal (2011)]{S11}
Sokolovsky, K.~V., Kovalev, Y.~Y., Pushkarev, A.~B., 
\& Lobanov, A.~P. 2011, \aap, 532, A38
\bibitem[Tchekhovskoy \etal (2008)]{T08}
Tchekhovskoy, A., McKinney, J.~C., \& Narayan, R. 2008, \mnras, 388, 551
\bibitem[Tchekhovskoy \etal (2009)]{T09}
Tchekhovskoy, A., McKinney, J.~C., \& Narayan, R. 2009, \mnras, 699, 1789
\bibitem[Uchida \& Shibata (1985)]{US85}
Uchida, Y., \& Shibata, K. 1985, \pasj, 37, 515
\bibitem[Uchida \& Shibata (1986)]{US86}
Uchida, Y., \& Shibata, K. 1986, Can. J. Phys., 64, 507
\bibitem[Vlahakis \etal (2000)]{VTST00}
Vlahakis, N., Tsinganos, K., Sauty, C., \& Trussoni, E. 
2000, \mnras, 318, 417
\bibitem[Vlahakis \& K\"{o}nigl (2003)]{VK03a}
Vlahakis, N. \& K\"{o}nigl, A. 2003a, \apj, 596, 1080
\bibitem[Vlahakis (2004)]{V04}
Vlahakis, N. 2004, \apj, 600, 324
\bibitem[Wang \& Zhou (2009)]{WZ09}
Wang, C.-C., \& Zhou, H.-Y. 2009, \mnras, 395, 301
\bibitem[Wilson \& Yang (2002)]{WY02}
Wilson, A.~S., \& Yang, Y. 2002, \apj, 568, 133
\bibitem[Wong \etal (2011)]{W11}
Wong, K.-W., \etal 2011, 736, L23
\bibitem[Young et al. (2002)]{YWM02}
Young, A.~J., Wilson, A.~S., \& Mundell, C.~G. 2002, \apj, 579, 560
\bibitem[Yuan \etal (2012a)]{Y12a}
Yuan, F., Wu, M., \& Bu, D. 2012, \apj, 2012, 761, 129
\bibitem[Yuan \etal (2012b)]{Y12b}
Yuan, F., Bu, D., \& Wu, M. 2012, \apj, 2012, 761, 130
\bibitem[Zakamska \etal (2008)]{Z08}
Zakamska, N.~L., Begelman, M.~C., \& Blandford, R.~D. 2008, \apj, 679, 990
\end{thebibliography}
\end{document}